\documentclass[10pt]{article}
\usepackage[letterpaper]{geometry}
\usepackage{hicss}
\usepackage{times}
\usepackage[none]{hyphenat}
\usepackage{url}
\usepackage{latexsym}
\usepackage{minted}
\usepackage{indentfirst}
\usepackage{graphicx}
\graphicspath{{images/}}
\usepackage[
    style=apa,
  ]{biblatex}

\usepackage{tabularx}
\usepackage{array,ragged2e} 
\usepackage{hyperref} 

\usepackage{enumitem}

\newcolumntype{P}[1]{>{\RaggedRight\arraybackslash}p{#1}}

 \usepackage[normalem]{ulem}
 \useunder{\uline}{\ul}{}

 \usepackage{lipsum} 
\usepackage{tabularx} 
\usepackage{ragged2e} 

\usepackage{quoting} 
\usepackage{adjustbox}

\usepackage{multirow}
\usepackage{booktabs}
\usepackage{tabularx}
\usepackage{xcolor}
\graphicspath{ {./images/} }

\title{Developer Productivity With and Without GitHub Copilot: A Longitudinal Mixed-Methods Case Study}

\author{
\begin{tabular}[t]{c}
Viktoria Stray\\
University of Oslo, SINTEF \\
\underline{stray@ifi.uio.no}
\end{tabular}
\And
\begin{tabular}[t]{c}
Elias Goldmann Brandtzæg\\
University of Oslo \\
\underline{eliasgb@ifi.uio.no}
\end{tabular}
\And
\begin{tabular}[t]{c}
Viggo Tellefsen Wivestad\\
SINTEF\\
\underline{viggo.wivestad@sintef.no}
\end{tabular}
\AND
\makebox[\textwidth][c]{%
  \begin{tabular}[t]{c}
  Astri Barbala\\
  SINTEF\\
  \underline{astri.barbala@sintef.no}
  \end{tabular}
  \hspace{4em}
  \begin{tabular}[t]{c}
  Nils Brede Moe\\
  SINTEF\\
  \underline{nils.b.moe@sintef.no}
  \end{tabular}
}
}
 
\date{}

\begin{document}
\maketitle
\begin{abstract}



This study investigates the real-world impact of the generative AI (GenAI) tool GitHub Copilot on developer activity and perceived productivity. We conducted a mixed-methods case study in NAV IT, a large public sector agile organization. We analyzed 26,317 unique non-merge commits from 703 of NAV IT's GitHub repositories over a two-year period, focusing on commit-based activity metrics from 25 Copilot users and 14 non-users. The analysis was complemented by survey responses on their roles and perceived productivity, as well as 13 interviews. Our analysis of activity metrics revealed that individuals who used Copilot were consistently more active than non-users, even prior to Copilot’s introduction. We did not find any statistically significant changes in commit-based activity for Copilot users after they adopted the tool, although minor increases were observed. This suggests a discrepancy between changes in commit-based metrics and the subjective experience of productivity.

\end{abstract}

\subsubsection*{Keywords:}

Agile software development, AI tools, Team collaboration,  Empirical software engineering, Developer experience, AI-assisted programming


\section{Introduction}
\label{sec:intro}
GenAI, particularly tools like ChatGPT and GitHub Copilot, have been rapidly adopted due to their promised ability to improve productivity and streamline workflows \parencite{Nguyen2023Generative, brynjolfsson2025generative}. GenAI tools assist developers by providing intelligent code suggestions, automating repetitive tasks, enhancing problem-solving capabilities \parencite{imai2022github,dohmke_github_2023}, and improved coding practices  \parencite{Zhang_2023}. However, while the GenAI technology seems appealing because of its promised productivity, there is a lack of empirical evidence on productivity from industrial studies.

Researchers and practitioners have been interested in the productivity of software engineers for many decades. Yet, the questions of how to measure or even define developer productivity remain elusive \parencite{forsgren2021space}. While productivity is often defined as output per unit of time \parencite{meyer2014software}, traditional metrics such as lines of code, number of tasks, or bugs fixed have been criticized for lacking validity and context \parencite{ralph2018construct, petersen2011measuring}. Developer productivity is inherently complex and multidimensional, shaped not only by tools and tasks, but also by team dynamics and the work environment \parencite{forsgren2021space}.

The emergence of GenAI tools such as GitHub Copilot and ChatGPT has further complicated the picture. Developers now spend less time typing code and more time reviewing AI-generated code suggestions \parencite{stray2025generative}. This shift renders many traditional productivity metrics even less meaningful. Furthermore, GenAI tools often lack the domain-specific knowledge required to address context-sensitive challenges, meaning that developers still rely on human collaboration for effective problem-solving when developing large systems \parencite{stray2025generative}. 

Although many claim that GenAI will boost productivity, clear evidence is still lacking. Some of the studies on developer productivity and GenAI use apply controlled lab settings \parencite{AlAhmad2024} or controlled field experiments \parencite{cui2024effects}. These are great for studying causal effects, but often fail to encapsulate the complex and nuanced process of a real-world software development setting \parencite{brynjolfsson2025generative}. Other studies rely on self-reported data or interviews from an industrial setting \parencite{Wivestad2024island}. However, conclusions based on self-reported data also have their limitations. As a result, there is a general acknowledgment that evaluating impacts on productivity in a real-world setting requires a more nuanced approach, one that combines both quantitative and qualitative methods to capture changes in developer experience, perceived effectiveness, and workflow dynamics \parencite{forsgren2021space}, and specifically how objective developer activity measures relates to developers’ perceived productivity \parencite{Wivestad2024space}.

Motivated by the need to understand in what way GitHub Copilot affects productivity in a real-world setting, we ask the following research questions:
\begin{itemize}
    \item  \textit{RQ1: How does Copilot adoption influence GitHub activity?}
    \item \textit{RQ2: How does perceived productivity relate to GitHub commit activity among Copilot users?}
\end{itemize}

We investigate changes in productivity, using a mix of usage data and developers’ own reflections by combining survey responses, interview insights, and analysis of GitHub activity over a two-year period. We make our analysis scripts publicly available, offering a reusable foundation for future research on developer activity using GitHub data (see Appendix A).

\section{Background}
\label{sec:back}
\subsection{Developer productivity}
Software development is a rapidly evolving field where systems must be continually adapted to new technologies and shifting requirements, which makes productivity a critical factor \parencite{wagner2019defining}. 
Productivity is generally defined as "\textit{the ratio of output divided by input}" \parencite[p.318]{petersen2011measuring}. Examples of \textit{outputs} include "\textit{quality and quantity in terms of functions, lines of code, implemented changes}" \parencite[p.318]{petersen2011measuring}, while \textit{inputs} are the efforts required to create these outputs \parencite{petersen2011measuring}. Measuring productivity can help us understand how to make software development more efficient, whether by reducing cost, improving quality, or speeding up the process \parencite{scacchi1995understanding}. Understanding these input and output factors gives companies better control over what to improve \parencite{wagner2018systematic}, although doing so is not always straightforward.

\label{subsec:how_to_measure_prod}
Throughout the history of software development, a variety of methods have been used to measure productivity, and there is broad agreement that this is challenging. \textcite{wagner2019defining} argues that measuring the input (cost) is relatively straightforward, but defining the output (software quantity and quality) is more difficult. \textcite{wagner2018systematic} provides a historical overview of how the different approaches for measuring productivity have evolved over time. Early productivity measurements primarily focused on technical factors, such as \textit{Lines of Code} (LOC), \textit{Function Points} (FP), and code complexity \parencite{wagner2018systematic}. Over time, researchers and practitioners recognized the limitations of purely technical metrics and began exploring the impact of non-technical, or "soft" factors, such as communication, collaboration, and team dynamics, on developer productivity \parencite{wagner2018systematic}. \textcite{murphy2019predicts} suggests that productivity measurement techniques can be broadly divided into two categories: \textit {objective metrics}, such as lines of code written per week, and \textit{subjective metrics}, like self-assessments and peer evaluations in surveys \parencite[p.583-584]{murphy2019predicts}. Regardless of multiple approaches, no single metric fully captures productivity in software development \parencite{jaspan2019no}.

For example,  the LOC metric is the amount of code a developer produces in a unit of time \parencite{scacchi1995understanding}. This metric is a limited measure of productivity because it does not account for code complexity \parencite{scacchi1995understanding}, different programming languages (high-level languages like Python may have a lower LOC compared to low-level languages like C\#) \parencite{petersen2011measuring}, or variations in individual developers' skills (some developers may produce the same functionality with fewer LOC) \parencite{petersen2011measuring}. A high LOC count can create the impression of productivity without actually achieving it \parencite{blackburn1996improving}. On the other hand, subjective methods like self-reported ratings and peer evaluations capture different aspects of a developer’s work. These approaches can reveal factors that influence individual productivity \parencite{murphy2019predicts}. Still, such methods also have their limitations. The results may be affected by cognitive biases \parencite{murphy2019predicts} and people might respond in ways that portray themselves favorably due to \textit{social-desirability bias} \parencite[p.2]{rosenman2011measuring}. Because developer productivity involves not only writing code but also problem-solving, collaboration, and bug fixing, a single metric inevitably oversimplifies the phenomenon. Therefore, researchers suggest designing a set of metrics that can provide a more complete view \parencite{jaspan2019no}.

\subsection{GenAI's effect on productivity}
In the annual developer survey conducted by StackOverflow in 2024, 76\% of the 60,907 respondents report that they are either using or planning to use AI tools for development \parencite{stackoverflow2024}. The share of current users increased from 44\% to 62\% in 2024. This illustrates the widespread adoption of AI tools in software development, with developers increasingly integrating solutions like ChatGPT and GitHub Copilot into their workflows. A key driver of this trend is increased productivity, with 81\% of surveyed developers citing it as a primary benefit.

Several recent studies indicate that using GenAI may transform software development by offering tools that boost productivity and optimize workflows in multiple ways \parencite{strayetal_ist_editorial}. The impact of GenAI extends across multiple development aspects and appears to accelerate the transformation of traditional roles, tools, and processes in software development \parencite{Sauvola2024}. These tools enhance productivity by automating repetitive software engineering tasks \parencite{ulfsnes2024transforming}, allowing developers to focus on complex problems \parencite{stray2025generative}. They assist developers in code generation, completing code snippets, increasing test coverage, answering questions about the code, and working with legacy code by explaining or translating it into modern languages \parencite{Ebert2023}. Additionally, these tools have been shown to be promising to aid in error detection, accelerate prototyping efforts \parencite{AlAhmad2024}, and reduce the time spent searching for information, which enables developers to maintain a sense of “continuous progress” \parencite{stray2025generative}. Moreover, they may lower skill barriers between different phases of software development, which can further accelerate the overall development process \parencite{rajbhoj2024accelerating}.


Employing a mixed-methods approach in a corporate setting, \textcite{brazil} analyzed observations, surveys, and company development data, finding that most participants felt they completed tasks more quickly when using GenAI tools. In the study by \textcite{Peng2023}, developers were asked to implement an HTTP server in JavaScript as quickly as possible. Those with access to GitHub Copilot completed the task 55.8\% faster than the control group. Notably, the productivity gains were larger for less experienced developers. 
Similarly, \textcite{cui2024effects} studied the impact of Copilot in real workplace settings across three companies, including Microsoft and Accenture. They used metrics such as pull requests, commits, and builds as proxies for completed tasks, and compared developers who were randomly assigned Copilot to a control group. The results showed a 26.08\% increase in completed tasks for the Copilot group, with the greatest gains seen among less experienced developers. Although the authors describe their study as a controlled field experiment, it differs from lab-based experiments because it takes place in real work settings, making the results more relevant to everyday software development.

\textcite{humlum2025large} reported similarly modest productivity effects from GenAI tools in a large-scale survey of over 25,000 employees across 7,000 workplaces and 11 different occupations in Denmark, including software developers. On average, workers using GenAI tools saved only 3\% of their time. The study found that productivity gains varied by occupation, with software developers reporting greater benefits than, for example, teachers. The results also depended on how actively organizations supported GenAI adoption, for instance, through access of internal models or tutorials on how to use such tools. Among software developers, those who were encouraged to use GenAI saved 6.5\% of their total work hours and used the tools on 57\% of workdays. In contrast, developers who were not encouraged reported only 3.9\% in time savings and 38\% tool usage. Despite these small gains, the study found no significant effects on broader economic outcomes, such as total hours worked or wage increases. 
In short, controlled task studies report larger productivity effects than field studies and workplace surveys \parencite{Peng2023, cui2024effects, humlum2025large}.


Although the adoption of GenAI tools in software development potentially can improve productivity, \textcite{stray2025generative} found that many developers had trust issues and spent additional time verifying the suggestions generated by AI tools because they encountered inaccuracies in the outputs. Further, \textcite{Simkute2024} identified a shift in the developer's role when using GenAI: instead of creating outputs, developers now focus on supervising and evaluating AI-generated results. This shift can be challenging, especially given the limited explainability of GenAI outputs. Similar concerns have been raised in the context of routine work automation, where one study found that while "boring" tasks were largely eliminated, employees' sense of meaningfulness was negatively affected due to growing insecurities around their professional role \parencite{staaby2021automation}.



\section{Method}
\label{sec:method}
 This case study was conducted in NAV IT, the technology division of a large Norwegian public sector organization. 
 With nearly one thousand employees, the division includes cross-functional teams of developers, designers, infrastructure and cloud specialists, architects, data scientists, and team leads. In September 2023, 100 software developers at NAV IT volunteered to receive a GitHub Copilot license. By May 2025, the number of Copilot users had grown to 250. 

We adopted a case study approach to understand how Copilot influenced development work in practice. Our study combines quantitative and qualitative data, including survey responses, interview insights, and analysis of the developers' activity on GitHub. 

In November and December 2023, we conducted 13 interviews with developers, each lasting between 30 and 60 minutes, with an average duration of 47 minutes. The interviews were transcribed using an AI tool, but we manually reviewed and corrected the transcriptions for accuracy and clarity while listening to the recordings. The transcripts were analyzed thematically, using a combination of open coding, memoing, and comparison. 

The survey was distributed between March and April 2024. As part of the survey, respondents who stated having adopted GitHub Copilot were given the following question: \textit{Since getting access to GitHub Copilot, have you noticed a change in the following: Your own productivity} (from “Major decrease” (-2) to “Major increase” (+2)), which we have used in this case study. While the survey was primarily designed to map out developers' attitudes regarding GenAI, the survey also asked if participants were willing to share their GitHub usernames so that we could link their answers to their GitHub activity. A total of 63 participants provided their username. 




\begin{figure}[b]
\vspace{-14pt}
  \centering
\includegraphics[width=\columnwidth]{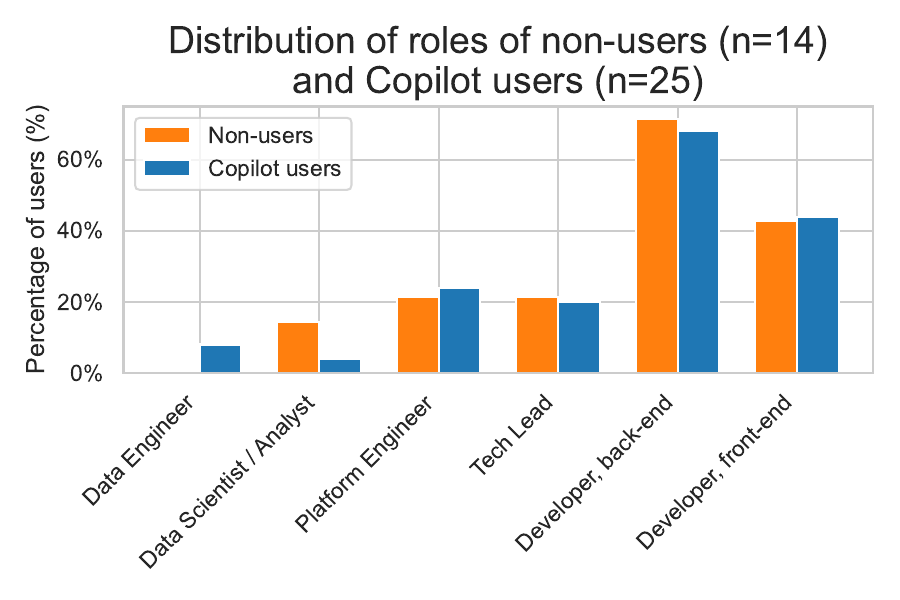}
  \caption{Self-reported roles of the 39 employees whose GitHub activity was analyzed.}
  \label{fig:roles}
\end{figure}

\begin{figure*}[tbh]

  \centering
  \includegraphics[width=0.9\textwidth]{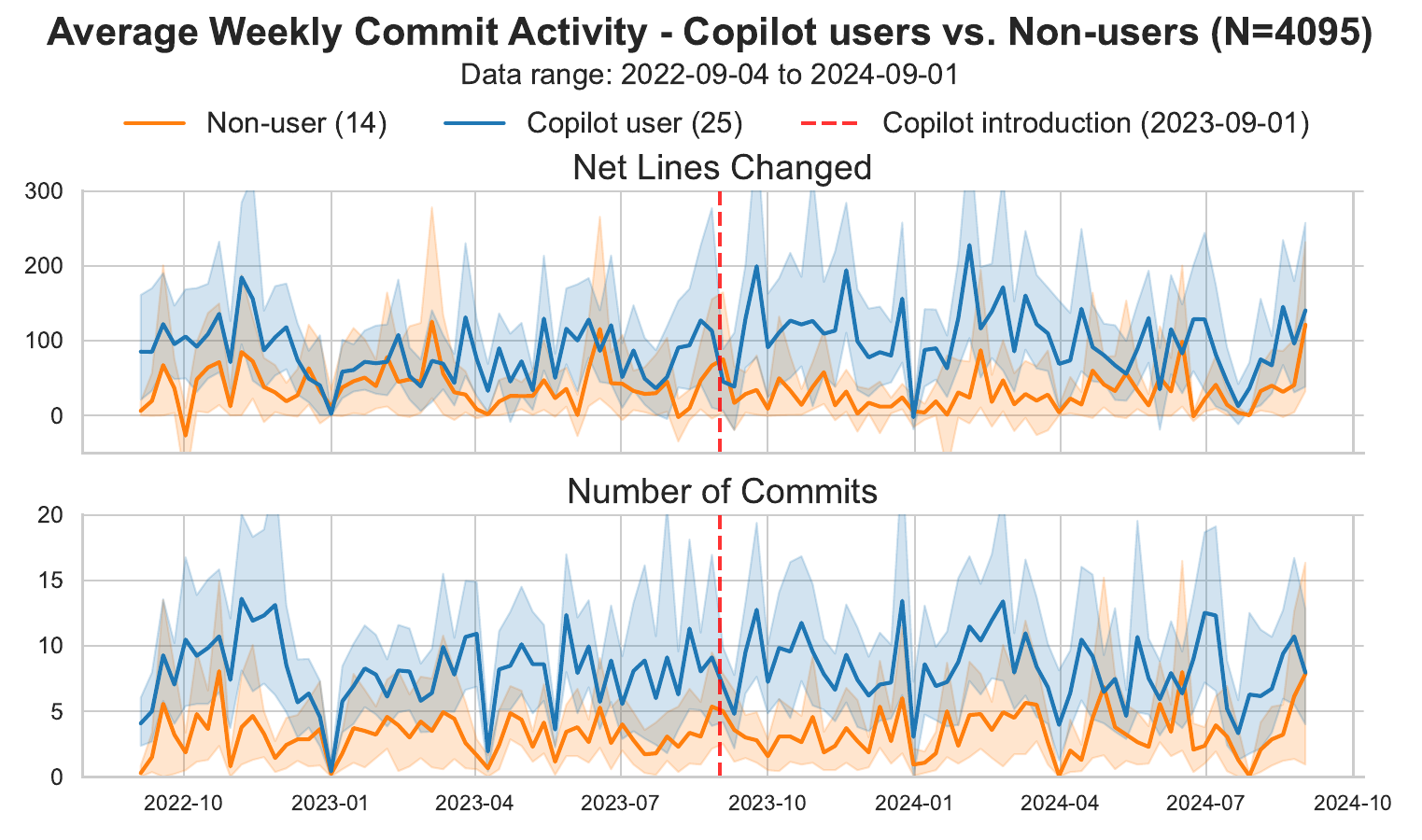}
  \caption{Time series showing the average weekly commit activity among GitHub users and non-users. The top plot shows net lines changed (lines added - lines removed), while the lower plot shows the average commit frequency (i.e., the average number of commits per week). The red vertical line shows when GitHub Copilot was introduced in the organization. The shaded areas around each line represent the 95\% confidence intervals.}
  \label{fig:lineplot_commits}
\end{figure*}
In order to retrieve the commit activity for the relevant developers, several custom Python scripts were developed. First, a list of all NAV IT's public repos was collected using GitHub's APIs. Next, using the Python framework PyDriller \parencite{PyDriller}, each repository was cloned, mined for data locally, and then deleted, storing only commit data for participants who provided their username in the survey.

The original dataset consisted of 37,974 commits. However, this dataset contained some issues that were discovered by a mix of statistical and manual inspection: \textbf{Duplicates}: Some commits were present across multiple repos and were removed. \textbf{Outliers}: Some commits contained an extreme amount of code insertions or deletions. Manual inspection revealed that this typically was due to non-human-generated code (e.g., someone adding or removing log or build files). The dataset was therefore trimmed at the 95th percentile, calculated individually for each commit activity metric. \textbf{Low activity users}: Users with fewer than one commit per week on average were excluded from the study, as they were not considered representative of professional developers.

The cleaned dataset consisted of 26,317 commits from 703 repositories spread across 39 developers: 25 Copilot users and 14 non-users. We include the roles of these participants as stated in their survey responses; see Figure \ref{fig:roles}.

Individual commit data can vary significantly in both size and frequency. To better investigate the overall trends for a broad concept like productivity, we aggregated the data into weekly chunks. This ensures that a range of contribution styles would be considered as "productive", both small but frequent contributions, and few but large contributions. For weeks with no commits, we manually imputed the value "0". The final dataset consisted of N=4095 weekly observations, based on 105
weeks of activity data for each of the 39 developers.



The data spans a two-year period, allowing us to compare developer activity before and after the introduction of Copilot. Most commits came from the organization's two main GitHub domains. 
Some findings from the surveys and interviews have been reported previously; however, we have conducted further analysis on this data to explore our new findings on GitHub activity \parencite{Wivestad2024island, Wivestad2024space,strayFSE,barbala2025generative, Brandtzaeg2025CopilotNAV}.

\section{Results}
\label{sec:results}
\begin{figure*}[thb]
  \centering
  \includegraphics[width=0.9\textwidth]{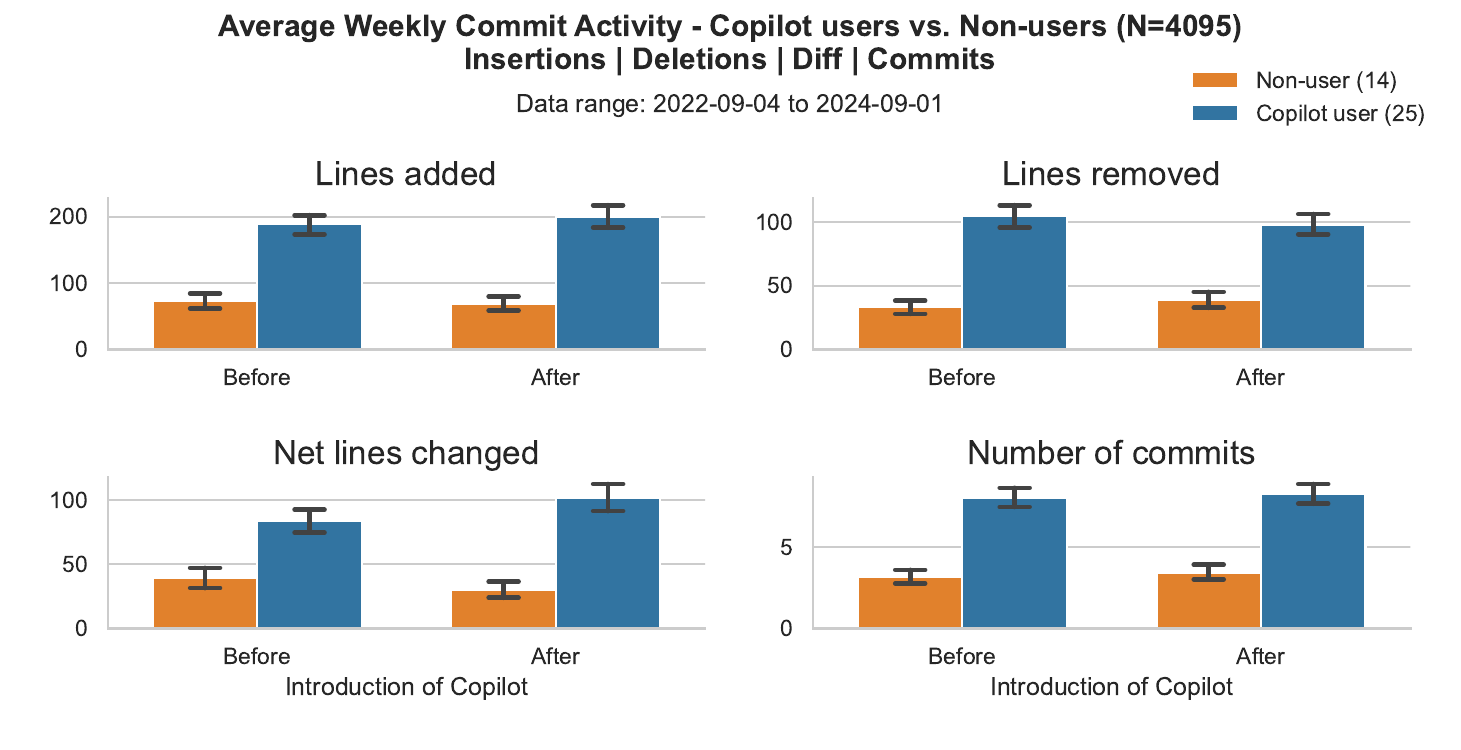}
  \caption{Average weekly commit contributions for non-users and Copilot users for the periods before and after Copilot adoption. The error bars show the 95\% confidence interval for all the weeks in each period.}
  \label{fig:barplot_group1}
\end{figure*}

Not all developers jumped on GitHub Copilot immediately. In NAV IT, about 100 developers volunteered for early access, but others held back. Some were reluctant or skeptical developers who either declined to adopt Copilot or tried it and stopped. Some seasoned developers did not perceive sufficient value to justify a new tool. One developer (I11) described that his team had already eliminated a lot of boilerplate through architectural choices, so Copilot felt superfluous to them. Another developer was openly unconvinced after seeing Copilot in action and chose not to opt in to get a license: \textit{"I think I might want to see examples of it working well before I… try it out myself. We had a little demo, and I didn't think it worked well at all… So I wasn't that impressed. […] So I don't need to spend time on it right now"} (I9).
\subsection{Developer Activity: Copilot Users vs. Non-users}\label{subsec:1}




%

When comparing one year of commit data before and after the introduction of Copilot for 39 employees in NAV IT, we found a clear pattern: Developers who eventually adopted Copilot (n=25) were already pushing commits more frequently and contributing more code overall per week than their colleagues who had not adopted Copilot (n=14), as shown in Figure \ref{fig:lineplot_commits}. This gap was evident both in the year before Copilot was available and continued in the year after. On average, Copilot users committed more than twice as frequently as non-users and made greater weekly contributions. The official introduction of Copilot did not produce a noticeable spike in the activity for either group (seasonal fluctuations, such as dips around Christmas and summer breaks, are visible for both groups).


The Mann–Whitney U tests comparing Copilot users and non-users also confirmed that Copilot users were significantly more active than non-users ($p < 0.00555$), both before and after adoption. In other words, Copilot adopters were already significantly more active developers to begin with, and they remained more active afterwards, indicating a self-selection effect.



However, the quantitative data does not capture fully what lies behind these findings, and our interviews revealed a slightly more nuanced perspective. For example, Interviewee 5 stated, \textit{"I don't think the tool is the most important part of the job. Understanding the task and collaborating are more important."} This thus suggests that the social and cognitive factors of the job are also crucial elements in determining developer practices and productivity.


To dig deeper into the underlying factors behind the findings presented above, we examined the volume of code changes (lines of code added and removed) for each group, before and after adoption. 
Figure~\ref{fig:barplot_group1} aggregates the weekly activities into two periods: before and after Copilot’s introduction. The figures show the average weekly code additions, deletions, and net difference, as well as the average number of weekly commits, for non-users (orange) and Copilot users (blue).
The bar charts reinforce the earlier trend: Copilot users consistently produce a higher volume of code changes per week than non-users. Before Copilot’s launch, users were adding ~188 lines and deleting ~105 lines per week on average, compared to non-users’ ~80 additions and ~40 deletions. After Copilot’s launch, the user group’s insertions ticked up slightly (to ~200/week) while deletions ticked down (~98/week), yielding an increase in net lines of code (+16 net lines as average weekly). Non-users, by contrast, showed a small decline in both additions 
(~70) and deletions (~30) in the same period (thus a roughly flat net change). These shifts were relatively small.

This suggests that Copilot might have enabled users to slightly increase their net code contribution by adding slightly more and deleting slightly less, while maintaining their average commit frequency. Interviewee I13, a Copilot user, described this effect as generating more initial code that could be refined later: \textit{ “Just getting code generated as a starting point to build on is incredibly good}” (I13). 


It is important to note that we found no evidence of any negative impact on code quality metrics from Copilot adoption. The structural metrics (e.g., function complexity, average module size) remained virtually unchanged and showed no significant differences between users and non-users. In other words, the Copilot group’s code was not measurably “worse” (or better) in complexity or size due to using the AI assistant. Interview insights support this: none of the developers felt Copilot degraded the quality of their code – if anything, the concerns were about hidden mistakes or subtle bugs, not systematic complexity.

\subsection{Perceived Productivity vs. Measured Output among Copilot users}\label{subsec:2}

\begin{figure*}[t]
  \centering
  \includegraphics[width=\textwidth]{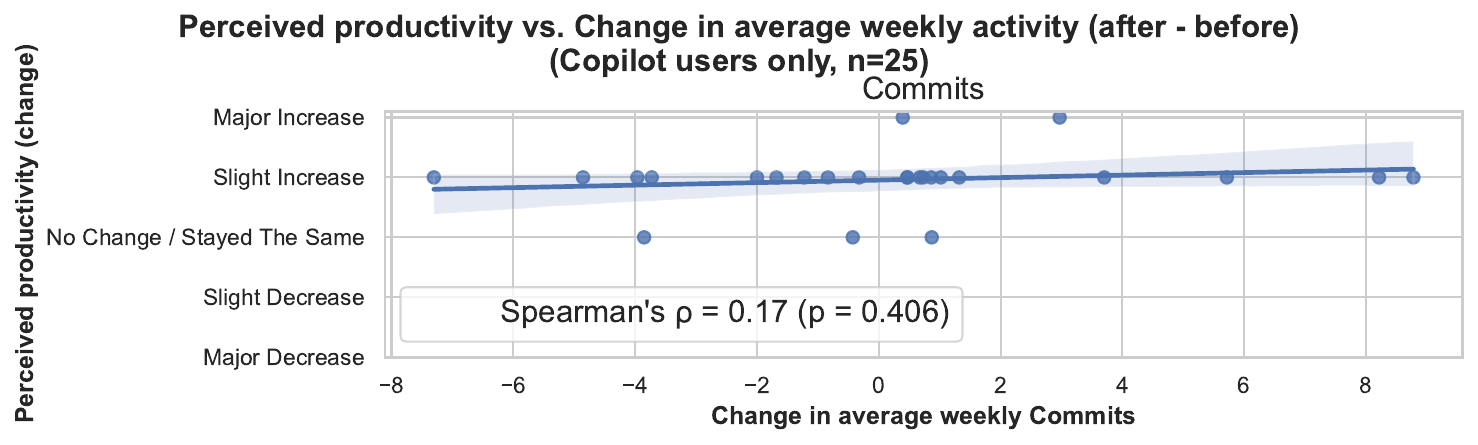}
  \caption{Correlations between change in commits and perceived productivity. Y-axis represents the 5-point Likert scale responses for perceived change in productivity. X-axis shows the change in average weekly activity.}
  \label{fig:scatterplot_commits}
\end{figure*}

Our quantitative analysis found no strong correlation between a developer's change in weekly commits and their self-reported productivity change from the survey. Figure~\ref{fig:scatterplot_commits} illustrates this for commit counts. Although a slight upward trend is visible, the correlation is not statistically significant (Spearman $\rho \approx 0.17$, $p = 0.40$).

Most points in Figure~\ref{fig:scatterplot_commits} cluster around ``No change'' to ``Slight Increase'' in perceived productivity, spanning a range of actual commit changes (some users actually committed less on average but still felt a bit more productive, and vice versa). Quantitatively, across all six activity metrics we examined, the correlations with perceived productivity were not found to be significant. The strongest correlation was observed for commit count, with a $\rho = 0.17$, which is considered a weak correlation, but with a p-value of 0.406, which did not allow us to reject the null hypothesis when using a Bonferroni corrected threshold for six tests of $\alpha=0.0083$. In practical terms, most developers seemed to feel a bit more productive, though no clear relationship could be established between this and writing more code. Developers who felt ``slightly more productive'' might attribute that to things like quicker completion of certain tasks or reduced frustration, which do not necessarily show up as big jumps in commit frequency. Another explanation could be that the perceived increase in productivity is partly a result of a subject-expectancy effect, where the introduction of a highly anticipated tool creates a subjective sense of improved performance without a measurable productivity increase. As one interviewee noted: ``\textit{I have no faith that we'll get good numbers (to measure this)}'' (I1), expressing skepticism that productivity gains from AI would be captured by metrics like lines of code.



Interestingly, the one activity metric that showed a negligible correlation with perceived productivity was net lines of code ($\Delta$Diff, $\rho \approx 0.09$, where  $\Delta$Diff is average weekly net diff after adoption minus average weekly net diff before adoption). This indicates that increases in code output volume had essentially no relationship 
with the feeling of being more productive. Meanwhile, a few who reported “major increase” in productivity did not stand out in code metrics; their boost may have come from qualitative improvements like less context-switching or more mental energy for challenging parts of the work. One senior developer reflected on this, saying that Copilot did not make hard problems go away, but it did save him time on template code and searching for syntax, which made his day “flow” better (I12). 
In the interviews, many Copilot users stated to feel that they felt faster and more efficient in their day-to-day work. Interviewee I12 stated, \textit{"I'm more productive, absolutely, since I started using Copilot. I get a solution to try faster, and it often works. Sometimes it's wrong... but regardless, it's rarer now that I sit and struggle with something all day that might not lead to anything."}
Our findings suggest that Copilot was mostly used to handle the more routine or repetitive parts of coding. At the same time, the interviewees stated that using the tool sometimes made the work more enjoyable, even fun, which may have also contributed to developers feeling more productive. One developer, I11,
explained how adopting GenAI tools had been a collective endeavour in his department due to certain eager colleagues: \textit{"I've been working with one person who is very enthusiastic [about GenAI]. So I thought I'd give it a shot, and that's why I signed up [to get the licence], actually, because he was so enthusiastic"}. The colleague in question, I8, was pleasantly surprised by how the new tools had added an element of playfulness to his day: \textit{"I haven't played with Copilot as much as I would like [at work], but I've been using it in my spare time [...]It's fun when part of the programming becomes a hobby, you know?"} This suggests that social and cognitive factors of GenAI use also directly impact the perceived productivity of software developers, an aspect that is often overlooked in studies focusing solely on quantitative metrics. 




\section{Discussion}
\label{sec:discussion}
We will now discuss our findings in light of our research questions. 
To answer our first question (\textit{How does Copilot adoption influence GitHub activity?}), we analyzed and compared developer activity for Copilot users and non-users by using commit-based activity metrics over a two-year period.





For Copilot users, our findings suggest a relatively small increase in average weekly activity; however, there is no statistically significant difference after adopting Copilot, which differs from prior studies \parencite{Peng2023, rajbhoj2024accelerating, AlAhmad2024, cui2024effects}. For example, \textcite{Peng2023} found that participants with access to GitHub Copilot completed standardized programming tasks 55.8\% faster. Similarly, \textcite{rajbhoj2024accelerating} studied the development of a small business application and reported a 70.66\% productivity gain, primarily due to ChatGPT’s ability to generate structural code. 

However, a major limitation of these studies is that they do not evaluate long-term effects or usage of GenAI tools in real, complex organizational environments. Instead, they focus on short-term gains in controlled or semi-controlled settings, either by measuring task completion time \parencite{Peng2023}, or by simulating software projects to illustrate how GenAI tools can assist developers during different phases of the development process \parencite{AlAhmad2024, rajbhoj2024accelerating}. 

Our findings align with those of \textcite{humlum2025large}
, who report modest productivity gains from GenAI tools in a large-scale study of workers in Denmark. On average, users in the latter study saved only 3\% of their time. \textcite{humlum2025large} argue that the lack of significant effects may be because we are still in an early phase of adoption, where organizations are adjusting workflows and making long-term investments. Following the results from this study, we may observe the use of GitHub Copilot too early to detect clear changes in activity or productivity. This is also similar to the findings of \textcite{brynjolfsson2025generative}, reporting relatively small productivity gains from GenAI tools in the customer support domain.


When analyzing Copilot adopters vs. non-adopters, one interesting finding is that Copilot’s adopters were among the more active developers even before the tool’s introduction. As illustrated in Figure~\ref{fig:barplot_group1}, Copilot users consistently displayed higher activity levels than non-users. This pattern exists both before and after Copilot adoption, suggesting that the users who chose to adopt Copilot were already more active in their development work. Possible explanations include differences in roles, personal traits such as curiosity, a stronger focus on frequent code production, or an interest in experimenting with new tools. It suggests a self-selection bias that those who write code more frequently (and perhaps are more enthusiastic about new tech) are the ones who decided to start using Copilot.

The reported roles of the two groups were quite similar, with many reported having multiple roles, and both groups containing managerial roles such as "Tech lead" (see Figure \ref{fig:roles}). 
This may explain why some individuals appear less productive in terms of commit-based metrics, as many of their responsibilities are likely less focused on writing code. 
For example, \textcite{cui2024effects} found that senior developers typically code less than junior developers, which explained much of the variation in software output. 

When answering our second research question (\textit{How does perceived productivity relate to GitHub commit activity among Copilot users?}), we
found a negligible correlation between perceived productivity and commit metrics. Our findings support that perceived productivity is multidimensional and personal. Developers in our study largely felt Copilot made them a bit more productive, primarily by alleviating drudgery and providing mental relief. This manifested in comments about being able to focus on “creative” tasks or having more fun coding. This discrepancy suggests that developers define productivity in different ways. While some may associate productivity with writing more code or making frequent commits, others may view it in terms of task completion, efficiency, learning, or collaboration. This is part of the reason why measuring productivity in knowledge work like software development, as \textcite{cui2024effects} puts it, "\textit{is notoriously difficult}" (p.6)
.
Instead, the value might lie in subtler forms of efficiency: time not spent on web searches, fewer context switches, or simply a smoother flow during coding.


One key challenge, as noted by \textcite{jaspan2019no}, 
is that no single metric can fully capture developer productivity, which depends on many interrelated technical and non-technical factors. This can explain why we found that increased perceived productivity does not necessarily align with measurable increases in commit-based activity. Subjective experiences of productivity are influenced by a broader set of factors, which may not be fully captured through metrics such as insertions, deletions, or commits. \textcite{meyer2017work} emphasize this in their study, concluding that perceived productivity is a highly personal concept and that perceptions of what is considered productive vary across developers. 

This broader understanding of productivity may help explain why none of the Copilot users in our sample reported a decrease in productivity. All participants indicated either “No change” or more (slight/major increase) on the Likert scale, even though many showed a decline in average weekly activity levels during the two-year analysis period (see Figure \ref{fig:scatterplot_commits}). This highlights how perceptions of productivity are complex and deeply subjective, making them difficult to capture through objective metrics alone. Developers vary in how they work, where some deliver in larger batches, others contribute more steadily throughout the week, and some follow more flexible schedules. These individual work patterns can significantly influence how activity is reflected in the data, further complicating the relationship between actual contributions and how productivity is measured or perceived.



In our study, several Copilot users reported increased productivity despite lower activity levels (see Figure \ref{fig:scatterplot_commits}), possibly because they spent more time evaluating generated code, as also observed by \textcite{stray2025generative}. This suggests that the tool may reduce cognitive load, streamline routine tasks, and improve developer experience—benefits not easily captured by traditional metrics like commits or lines of code. As developers in our study explained, the flow became better even though the output did not increase. 



Improving developers' feeling of flow reinforces prior critiques of conventional productivity metrics and highlights the value of integrating subjective, experience-based data. The SPACE framework’s \parencite{forsgren2021space}  emphasis on satisfaction, flow, and collaboration becomes particularly relevant here. Developers primarily used Copilot for boilerplate code, documentation, and repetitive tasks—areas where perceived value can be high, even if output volume doesn't drastically change.

\subsection{Implication for Practice}
 In our case, the developers maintained their higher output advantage post-Copilot introduction. This has an implication for evaluating AI tools: if one naively compared Copilot users vs. non-users after rollout, one might wrongly conclude Copilot “caused” more output, whereas our longitudinal view shows how this gap pre-existed GitHub Copilot (Copilot users were already committing more). This quantitative finding aligns with the skepticism voiced by some of the interviewees that simple metrics could be misleading.

 For practitioners, this means that evaluating AI assistants should include developer feedback and not just output metrics. Companies might consider surveying their developers on whether the tool saves them time or mental effort, complemented by observing over longer time periods to observe cumulative effects (positive or negative). Also, our findings suggest that Copilot adopters were already among the more active and productive developers prior to adoption. From a managerial perspective, this could indicate that when developers ask for new tools, it is often the most engaged and productive ones who do so. Supporting these requests may therefore serve more as a strategic investment in developer satisfaction and retention, rather than purely enhancing individual productivity.






\section{Conclusion and Future Work}
\label{sec:conclusion}

Our longitudinal study of GitHub Copilot adoption reveals a nuanced impact on developer productivity. In contrast to controlled experiments isolating Copilot’s effects, our real-world data showed no dramatic change in commit-based output after Copilot’s introduction. Notably, those who chose to use Copilot were already among the more active developers beforehand. Meanwhile, many developers perceived a productivity boost with Copilot, highlighting a persistent gap between measurable output and personal experience. This gap suggests that current metrics may fail to capture certain benefits of AI assistance, such as reduced mental load or smoother workflow. Going forward, the role of GenAI in productivity might be less about accelerating raw output and more about enabling developers to focus on higher-value tasks, minimize tedious effort, and sustain “flow” in their work. Future work should explore whether Copilot leads to more code being produced or just produced differently (and whether that is beneficial or problematic), and develop new ways to assess productivity that encompass quality, developer well-being, and long-term maintainability. As organizations continue to integrate AI coding assistants, our findings call for evolved productivity measures and further in-situ research to fully understand GenAI’s contributions to software development outcomes.

\section*{Acknowledgments}
We thank NAV IT, the interview participants, and the survey respondents for their valuable contributions. We are especially grateful to those who shared their GitHub username, enabling us to link their activity with their survey response. This work was supported by the Research Council of Norway through the projects TransformIT (grant 321477) and Kairos (grant 357147).




\label{sec:appendix}
\section*{Appendix A: Python Scripts}
Our scripts for mining GitHub data, cleaning, and
analyzing the dataset, performing statistical tests, and generating plots are available at: 
\href{https://figshare.com/s/736af5662435675e7914}{https://figshare.com/s/736af5662435675e7914}









\printbibliography

\end{document}